\documentclass[fleqn,usenatbib]{mnras}

\usepackage{newtxtext,newtxmath}

\usepackage[T1]{fontenc}

\DeclareRobustCommand{\VAN}[3]{#2}
\let\VANthebibliography\thebibliography
\def\thebibliography{\DeclareRobustCommand{\VAN}[3]{##3}\VANthebibliography}

\usepackage{graphicx}	
\usepackage{amsmath}	
\usepackage{lineno}

\newcommand{\msun}{$M_{\odot}$}
\newcommand{\lcdm}{$\rm{\Lambda}$-CDM}

\title[MW Missing Halo]{The Milky Way's Missing In-Situ Halo}

\author[Lucey et al.]{
Madeline Lucey,$^{1}$
Robyn Sanderson,$^{1}$
Alexander H.~Riley,$^{2}$
Andreia Carrillo,$^{3}$
Jenna Samuel,$^{4,5}$
\newauthor
and Emma Tasso$^{6}$
\\
$^{1}$Department of Physics \& Astronomy, University of Pennsylvania, 209 S 33rd Street, Philadelphia, PA 19104, USA\\
$^{2}$Lund Observatory, Division of Astrophysics, Department of Physics, Lund University, SE-221 00 Lund, Sweden \\
$^{3}$Department of Physics and Astronomy, Carleton College, 1 North College St., Northfield, MN 55057, USA \\
$^{4}$Department of Astronomy, The University of Texas at Austin, Austin, TX 78712, USA \\
$^{5}$Cosmic Frontier Center, The University of Texas at Austin, Austin, TX 78712, USA \\
$^{6}$Graduate Center, City University of New York, 365 5th Avenue, New York, NY 10016, USA 
}

\date{Accepted XXX. Received YYY; in original form ZZZ}

\pubyear{\the\year{}}

\begin{document}
\label{firstpage}
\pagerange{\pageref{firstpage}--\pageref{lastpage}}
\maketitle

\begin{abstract}
The Milky Way's stellar halo preserves a record of its mass assembly history, encoding accretion events in its structural properties. Among the most prominent of these features is a strong break in the halo density profile at $\approx$20–30 kpc, long attributed to the apocenter pile-up of stars from the Gaia-Enceladus/Sausage merger. However, whether this interpretation is consistent with state-of-the-art cosmological simulations remains unclear. In this work, we compare the Milky Way's measured stellar halo density profile between galactocentric radii of 1--100 kpc to those of Milky Way-mass galaxies from the FIRE-2 and Auriga cosmological zoom-in simulation suites, spanning a total of 24 simulated galaxies. We find that simulated halo profiles are significantly steeper than Milky Way measurements within $\approx$15 kpc, and that profile breaks are rare in simulations and never as strong as the Milky Way's. Of the three galaxies with statistically significant breaks, only one exhibits a break attributable to apocenter pile-up of an accreted merger remnant. Decomposing the simulated profiles, we find that the accreted halo profiles are broadly consistent with Milky Way measurements, while the in-situ halo dominates within $\approx$20 kpc and drives the discrepancy. These results raise a compelling question: where is the Milky Way's in-situ halo? The origin of this tension may reflect systematic biases in current halo measurements, an overproduction of early spheroidal star formation in cosmological simulations, or more fundamental differences in the underlying physics (e.g., dark matter) governing the assembly and structure of the inner Galaxy.
\end{abstract}




\section{Introduction}

The Milky Way (MW) offers a uniquely detailed case study for understanding galaxy formation and evolution. With billions of individually observable stars, its stellar populations carry detailed chemical and kinematic imprints of the Galaxy's formation history \citep{Grebel1999,Feast2000,Freeman2002,Sandage2006,Catchpole2016}. Its stellar halo, in particular, offers an especially sensitive testing ground for $\Lambda
$CDM and galaxy formation models.

Stellar halos preserve a record of their host galaxy's mass assembly history, built up through the accretion and disruption of satellites and star clusters \citep{White1978,Searle1978,Bullock2001,Bullock2005,Helmi2008,Jean-Baptiste2017,Monzon2026}. Their structural properties, including density profiles, morphology, and substructure, are sensitive to dark matter physics, including the halo mass spectrum and baryonic contraction \citep{Cooper2010,Pillepich2014,Monachesi2019,Lazar2020,Genina2023,Lucey2025,ForouharMoreno2025}. Characterizing these structural properties in the MW's stellar halo is therefore a critical step toward understanding the broader processes that govern galaxy assembly.

Decades of observational work have begun to characterize the structure of the MW's stellar halo. These studies find that the halo is flattened, with $q\approx0.6-0.7$ where $r=\sqrt{X^2+Y^2+(Z/q)^2}$  \citep{Juric2008, Bell2008,Deason2011,Iorio2018,Han2022,Lucey2026}, and most recent work additionally measures a tilt in the halo with respect to the Galactic disk plane \citep{Iorio2019,Han2022,Lucey2026}. There is also evidence that the flattening varies with Galactocentric radius \citep{Xue2015,Xu2018,Iorio2018,Hernitschek2018}, including a prolate ($q>1$) inner halo \citep[$r<10$ kpc;][]{Bowden2016,Posti2019,Lucey2026}. 

The density profile of the MW's stellar halo is most commonly described by a broken power law, with inner slopes in the range $\alpha\approx-1.5$ to $-3$, outer slopes of $\alpha\approx-4$, and break radii around $\approx20-30$ kpc \citep{Sesar2007,Watkins2009,Deason2011,Sesar2013,Faccioli2014,Zinn2014,Das2016,Hernitschek2018,Iorio2018,Chen2023,Lane2023}. Some works instead favor a doubly broken power law, with a second break at $\approx$10 kpc \citep{Han2022,Yang2022,Li2026}, and recent studies suggest the slope may vary along different lines of sight \citep{Lowing2015,Hernitschek2018,Amarante2024,Medina2024}. Measurements extending to the very distant outer halo ($r\gtrsim100$ kpc) recover similar slopes of $\alpha\approx-4$  \citep{Thomas2018,Medina2018,Stringer2021,Feng2024}, though some find an additional break at $r\approx160-210~\rm{kpc}$ \citep{Fukushima2018,Fukushima2019}. 

While the outer halo profile is now relatively well characterized, the radial density profile along the minor axis within 10 kpc remains poorly constrained. Characterizing this region is complicated by the presence of multiple overlapping components, including the halo, disk, and bulge, as well as complex substructure such as the Galactic bar and X-shaped feature \citep{Binney1991,Ness2016b,Debattista2017}.  However, existing measurements of the minor-axis stellar density distribution within the inner Galaxy show considerable disagreement, with power law slope estimates spanning $\alpha\approx-1.5$ to $-3$ \citep{Pietrukowicz2015,PerezVillegas2017,Iorio2018,Yang2022,Han2022,Lucey2026}. Key open questions are whether the outer halo, inner halo, and bulge constitute physically distinct components \citep{Lucey2021,Ardern-Arentsen2024,Lucey2026}, and whether the inner halo harbors a significant in-situ population separate from the accreted outer halo \citep{Carollo2010,Haywood2018,Di_Matteo2019,Naidu2020,Belokurov2022}. The in-situ inner halo may have formed through ELS-like (Eggen, Lynden-Bell and Sandage) dissipational contraction, in which early star formation occurred on halo-like random orbits \citep{Eggen1962}, or through the disruption of an old thick disk during a major merger \citep{Belokurov2020}.

Cosmological zoom-in simulations offer a powerful framework for interpreting these structural and formation questions, making predictions for the radial density profiles of stellar halos across a range of assembly histories. These simulations have halos built from a mix of dissipational collapse, accreted galaxies and disrupted disks \citep{Abadi2003,Zolotov2009,Elbadry2018,Monachesi2019,Yu2023,Tau2025}. Using the IllustrisTNG suite of simulations, \citet{Pillepich2018} found that, on average, MW-mass galaxies have a stellar density power law slope of $\approx$-4.3 for stars outside the half-light radius. Both the Auriga and FIRE-2 simulations produce stellar halo density profiles broadly consistent with a power law slope of $\alpha\approx-4$ \citep{Monachesi2019}, a result that holds even in the inner halo \citep[$r<10$ kpc;][]{Lucey2025}. 

Interestingly, observations of the MW consistently point toward shallower inner halo profiles than those predicted by cosmological simulations \citep{Sesar2007,Deason2011,Faccioli2014,Das2016,Iorio2018,Han2022,Yang2022,Lane2023,Li2026,Lucey2026}. However, when comparing the Auriga cosmological zoom-in simulations to the MW and nearby galaxies, \citet{Monachesi2019} find the simulated stellar halo profiles better match the observations when excluding in-situ stars. \citet{D'Souza2018} also found better agreement between the minor-axis surface brightness profiles of nearby galaxies and  MW-mass Illustris simulations when limiting to accreted stellar particles. While early numerical studies produced some broken profiles with shallower inner slopes \citep{Bullock2005,Cooper2010,Deason2013}, it remains to be seen whether these results hold in fully hydrodynamical zoom-in simulations. Comparing the accreted stellar profiles of semi-analytic models \citep{Bullock2005,Cooper2010} to Auriga simulations, \citet{Pu2025} find that Auriga profiles are more smooth. \citet{Font2020} fits broken power laws to simulated stellar halos for Milky Way-mass galaxies from the ARTEMIS cosmological hydrodynamic zoom-in simulations. They do not discuss whether single power laws are statistically preferred, but find the breaks tend to relate to the transition between in-situ dominated inner halo to accreted outer halo. In this work, we investigate further the potential discrepancy between MW measurements and cosmological predictions, including whether simulated halos exhibit comparable breaks in their density profiles.

This publication is laid out as follows. In Section \ref{sec:sims} we describe the Auriga and FIRE-2 simulations, while Section \ref{sec:MWLit} describes the MW literature used for comparison in this work.  We discuss the lack of profile breaks in the simulations in Section \ref{sec:break} and possible origins of the discrepancy with the MW literature in Section \ref{sec:accr}. Last, in Section \ref{sec:sum}, we summarize this work and the conclusions.

\section{Cosmological Zoom-in Simulations} \label{sec:sims}
This work draws on two complementary suites of publicly available cosmological zoom-in simulations of MW-mass galaxies, FIRE-2\footnote{\url{https://flathub.flatironinstitute.org/fire}} \citep{Wetzel2023} and  Auriga\footnote{\url{https://wwwmpa.mpa-garching.mpg.de/auriga/index.html}} \citep{Grand2024}. While the two suites share broadly similar numerical frameworks, they differ in key aspects of their approach to modeling baryonic physics. This makes them well-suited for examining how choices in subgrid modeling shape the resulting galaxy populations.

\subsection{FIRE-2}

From FIRE-2, we use the \textit{Latte} suite of seven isolated MW-mass galaxies \citep{Wetzel2016} and the \textit{ELVIS} suite of three Local Group-like pairs of MW-mass galaxies \citep{Garrison-Kimmel2019}. All simulations adopt the FIRE-2 physics model \citep{Hopkins2018b} and are run with the GIZMO\footnote{\url{http://www.tapir.caltech.edu/~phopkins/Site/GIZMO.html}} gravity and hydrodynamics code in meshless finite-mass (MFM) mode \citep{Hopkins2015}. We refer the reader to these papers for full implementation details, and summarize the key properties below.
All simulations assume a flat \lcdm\ cosmology with parameters consistent with \citet{Planck2014}. The \textit{Latte} suite (excluding m12w) adopts $\Omega_m$ = 0.272, $\Omega_b$ = 0.0455, $\sigma_8$ = 0.807, $n_s$ = 0.961, $h$ = 0.702. The \textit{ELVIS} pairs Thelma \& Louise and Romulus \& Remus use the cosmology from the original DMO \textit{ELVIS} suite: $\Omega_m$ = 0.266, $\Omega_b$ = 0.0449, $\sigma_8$ = 0.801, $n_s$ = 0.963, $h$ = 0.71.  The remaining pair, Romeo \& Juliet, along with m12w, instead adopt updated parameters from \citet{Planck2020}: $\Omega_m$ = 0.31, $\Omega_b$ = 0.048, $\sigma_8$ = 0.82, $n_s$ = 0.97, $h$ = 0.68.

Stellar feedback prescriptions are known to significantly influence the distribution of mass in simulated galaxies \citep{Pontzen2014,Lazar2020}. The FIRE-2 model includes stellar winds, radiation pressure from young stars, Type II and Type Ia supernovae, photoelectric heating, and photoionization, all of which collectively regulate star formation. The gas density threshold for star formation is $n_{SF}>1000~\rm{cm^{-3}}$, and feedback rates, luminosities, energies, and mass-loss rates are drawn directly from stellar evolution models \citep[STARBURST99;][]{Leitherer1999}. Across all 13 simulated galaxies, dark matter halo masses at $z=0$ fall in the range $\rm{M_{200c}= 0.7-1.3 \times 10^{12}~M_{\odot}}$ \citep{Samuel2020,Barry2023},  where  $\rm{M_{200c}}$ is defined as the mass enclosed within a sphere in which the average density is 200× the critical density of the universe. Initial stellar particle masses are 7070 \msun\ for the \textit{Latte} suite and 3500 \msun\ for the higher-resolution \textit{ELVIS} suite, with star particle softening lengths of $\approx$4 pc and dark matter force softening of $\approx$40 pc. The dark matter particle masses are 35,000 \msun\ for the \textit{Latte} suite and 19,000 \msun\ for the \textit{ELVIS} suite. This leads to numerical convergence on the $\approx$100 pc scale \citep{Wetzel2016}, sufficient to resolve the disk scale heights of $\lesssim$950 pc \citep{Yu2021}. Therefore, numerical effects should not strongly impact our study of the stellar halo which focuses on scales $>1$ kpc.

To understand the role of hierarchical formation in building the stellar halo  profile, we identify and track star particles that formed outside of the main progenitor galaxy across the simulation as a function of time. These star particles are formed in subhalos before they interact with the main branch progenitor. The redshift when the main progenitor reaches a mass that is 3 times larger than the next most massive luminous halo, $\rm{z_{MR_{3:1}}}$, is when the main progenitor emerges as the dominant host galaxy \citep{Santistevan2020,Horta2024}. After this redshift, the subhalos are tracked with the help of the ROCKSTAR halo catalogs and the halo merger trees \citep{Behroozi2013, Behroozi_etal_2013b}, along with every star particle that makes up a present-day substructure (satellites, streams, or phase-mixed) within this virial radius \citep{Panithanpaisal2021,Kundu2026}. We specifically use the catalogs from Tasso et al. in prep. which are updated to begin tracking as early as cosmic time = 1.26 Gyr, where $\leq$7.3\% of stars have formed across all simulations.

\subsection{Auriga}
The Auriga cosmological zoom-in simulations serve as a natural comparison to FIRE-2, sharing broadly similar resolution and MW-like disk morphologies at $z=0$ while differing in their baryonic physics prescriptions. We include six MW-mass galaxies at level 3 resolution, which most closely matches the FIRE-2 resolution used here, as well as five at level 4 resolution that were selected to host GES-like mergers \citep{Fattahi2019}. For a detailed description of the Auriga simulations we refer the reader to \citet{Grand2017} and \citet{Grand2024}, and summarize the key properties below.

The Auriga simulations adopt a flat \lcdm\ cosmology with parameters from \citet{Planck2014} and are run using the gravo-magnetohydrodynamics moving-mesh code AREPO \citep{Springel2010,Pakmor2016}. Host halos are selected from the dark-matter-only EAGLE simulations \citep{Schaye2015} as isolated MW-mass systems ($1<M_{200c}/[10^{12}~M_{\odot}]<2$ at $z=0$. The initial conditions for higher resolution zoom-in simulations are created using \textsc{Panphasia} \citep{Jenkins2013}. At level 3 resolution, dark matter particle masses are $3.6\times10^4$ \msun\ and stellar particle masses are $6.7\times10^3$ \msun, with softening lengths of $\approx$188 pc applied uniformly to gas, stars, and dark matter. Level 4 resolution runs have dark matter particle masses of $2.9\times10^5$ \msun\, stellar particle masses of $5.4\times10^4$ \msun, and softening lengths of $\approx$375 pc. Through resolution studies, \citet{Grand2017} finds numerical convergence for level 4 resolution on the $\approx$100 pc scale and resolves disk scale heights of $\gtrsim$200 pc. Therefore, these simulated galaxies should have sufficient resolution for our study. 

The Auriga simulations include primordial and metal-line cooling \citep{Vogelsberger2013}, and, similar to FIRE-2, a spatially-uniform redshift-dependent UV background for reionization \citep{FaucherGiguere2009}. The two simulation suites differ in several key aspects of their stellar feedback implementations. Both include Type II supernova feedback, but whereas FIRE-2 evolves supernova rates continuously with the aging stellar population, Auriga instead applies feedback instantaneously at the time of star formation \citep{Springel2003, Vogelsberger2013}. Auriga also lacks the early stellar feedback channels present in FIRE-2 — such as radiation pressure and photoionization — that are thought to regulate star formation on small scales. This is reflected in the star formation threshold, which is $n_{SF}>0.13~\rm{cm^{-3}}$ in Auriga compared to $n_{SF}>1000~\rm{cm^{-3}}$ in FIRE-2. Conversely, Auriga includes black hole seeding, accretion and feedback \citep{Springel2005_feedback,Marinacci2014,Grand2017} and magnetic fields \citep{Pakmor2017}, which are absent from the FIRE-2 simulations analysed here.

Halo catalogues are constructed using the SUBFIND halo finder \citep{Springel2001} and linked in postprocessing using the LHaloTree merger tree algorithm \citep{Springel2005_millenniumclustering}.
Based on these halo catalogues and merger trees, star particles are assigned to haloes at the time of their birth and connected to the $z=0$ snapshot information.
The resulting `particle lists' are used in this work to identify star particles as accreted/in-situ or to isolate stellar debris from a particular merger, while the merger trees are used to trace when satellites merge with the host.
We note that the halo catalogues, merger trees, and particle lists match those that are in the public data release \citep{Grand2024} and are discussed in further detail in Sections~2 and 3.1 of \citet{Riley2025}.
In this work, star particles are labelled as in-situ if they form within the main branch of the MW-mass host, otherwise they are labelled as accreted.

\subsection{Comparing FIRE-2 and Auriga}

The central MW-mass galaxies of both FIRE-2 and Auriga match the observed Stellar Mass--Halo Mass (SMHM) relation obtained from abundance matching, with the Auriga hosts lying toward the upper end but within the 1$\sigma$ scatter \citep{Behroozi2013c,Moster2013,Grand2017,Hopkins2018b}. The MW, by comparison, is a $\sim$1$\sigma$ outlier \citep{Licquia2015,Bland-Hawthorn2016,Cautun2020}. 
\textit{Latte} disks tend to form later than the MW disk and than disks in Auriga \citep{Belokurov2022, McCluskey2024}, which affects the timing of disk heating and in-situ halo formation.

Both Auriga \citep{Simpson2018_aurigasatellites, Grand2021} and FIRE-2 \citep{Wetzel2016,Garrison-Kimmel2019b,Samuel2020,Panithanpaisal2021,Cunningham2021} produce numbers of classical-mass ($M_\star > 10^5$~M$_\odot$) satellites broadly consistent with observations, including of the MW and M31.
Low-mass galaxy centrals in Auriga (Figure~2 of \citealp{Sales2022}) and FIRE-2 (Figure~B2 of \citealp{Feldmann2023}) generally lie above the SMHM relation inferred from abundance matching, though these constraints are subject to considerable modeling uncertainty \citep[see discussion in][]{Posti2019_smhm, Grand2024}.

Satellites in both suites are actively disrupting to form stellar streams \citep{Panithanpaisal2021, Riley2025}, with Auriga satellites producing more stellar streams than FIRE-2 \citep[e.g., Figure 12 of][]{Riley2025} and experiencing more stellar mass loss overall \citep{Riley2026b}.
Both suites produce more massive streams than observed in the MW, however the resulting tidal tails form at very low surface brightness ($\mu_V \gtrsim 32 \rm~mag~arcsec^{-2}$) and may not be detectable with current imaging \citep{Wang2017_tidal, Shipp2023}.
\citet{Shipp2023} show this the case for FIRE-2, while a similar analysis for Auriga has not been published.
The orbits of Auriga streams provide a closer match to the MW than FIRE-2 \citep{Shipp2023, Shipp2025}, though this result is subject to substantial host-to-host scatter \citep{Dropulic2025} and the MW and FIRE-2 disagreement may be alleviated with increasing numerical resolution \citep{Meziani2026}.

Relative to the observed stellar mass-metallicity relation \citep[e.g.,][]{Kirby2013}, Auriga satellites are metal-rich \citep{Grand2021,Kizhuprakkat2024,Riley2026} while FIRE-2 satellites are metal-poor \citep{Escala2018} at fixed stellar mass.
As accreted satellites build the stellar halo, these offsets propagate into the metallicity and possibly the mass of the accreted halo.
The FIRE-2 stellar halo mass fractions are consistent with MW estimates \citep{Sanderson2018}, and Auriga stellar halo masses and density profiles are broadly consistent with the GHOSTS sample \citep{Harmsen2017}, particularly when only considering accreted stars \citep{Monachesi2019}.
However, FIRE-2 inner halos are dominated by metal-rich in-situ stars \citep{Bonaca2017, Sanderson2018} and Auriga stellar halos are metal-rich compared to observational constraints \citep{Monachesi2019}\footnote{These comparisons for Auriga use all halo stars, not the old stars that trace the Galactic halo (Riley et al.~in prep).}.

\section{Measurements of the Milky Way's Stellar Halo Profile } \label{sec:MWLit}

The MW halo density profile has been the subject of extensive study, with a wide variety of stellar tracers and methodologies employed across the literature.   Distance uncertainties, selection effects and model selection complicate direct measurements and methods for correcting for these effects can vary. Many works use photometry and luminosity information of stars (e.g., standard candles) to reduce distance uncertainties \citep{Deason2011,Iorio2018,Han2022,Yang2022}, while others account for the uncertainties in their probabilistic models \citep{Kurbatov2024,Lucey2026}. Most recent work include detailed modeling of the selection function to account for completeness effects, including  crowding and dust extinction \citep{Iorio2018,Yang2022,Han2022,Kurbatov2024,Lucey2026}. Because different approaches introduce different biases and systematics, we summarize below a representative selection of results chosen to span this range. 

We divide the results into two sections based on the stellar tracer selection which fall into two categories: age-constrained populations (e.g., RR Lyrae, BHB stars) and chemically selected samples (e.g., metal-poor giants). Despite the differences in methodology and sample selection, there is reasonable agreement between the Milky Way results, especially at  $r>20$  kpc. For example, works using different tracers, different methodologies and different functional forms for the halo profile all find that the outer halo has a power-law slope of $\alpha \approx -3.99$ to $-4.6$ \citep{Watkins2009,Deason2011,Faccioli2014,Das2016,Hernitschek2018,Chen2023,Han2022,Yang2022,Lucey2026}. On the other hand, there is a wider range of slopes reported for the inner halo ($r\lesssim 20-30$ kpc), and differences in the exact radius at which the slope changes. We include works that span the full range of reported inner halo slopes from $\alpha = -1.5$ \citep{Yang2022} to $\alpha = -3.4 $ \citep{Kurbatov2024}.  For the sake of clarity in the figures, there are numerous recent works we do not include, but they give very consistent results to the profiles shown \citep[e.g.,][]{Chen2023,Lane2023,Amarante2024,Medina2024,Li2026}. The literature we focus on is chosen to not only span a range of methodologies and tracers, but also prioritize works that attempt to measure the halo at small Galactic radii ($r<5$ kpc).

\subsection{Age-constrained Tracers}

 In general, the following age-constrained tracers are primarily used for their standard-candle nature, which improves the distance inferences. They also tend to be old stars which helps to remove contamination from the disk population. However, this can also introduce biases if populations of different ages are not distributed equally throughout the halo. For example, if younger populations are more concentrated toward the Galactic center, works using only old stellar tracers would infer a shallower halo profile compared to the true total stellar mass profile. The following works may be impacted by such bias, but we also compare to chemically-selected sample, which will have their own distinct biases, to test this.  

\subsubsection{\citet{Deason2011}}

\citet{Deason2011} use a maximum likelihood method to model the stellar halo density profile using A-type stars selected photometrically from SDSS Data Release 8, a sample dominated by Blue Horizontal Branch (BHB) and Blue Straggler (BS) stars. BHB stars are predominately old ($\gtrsim$ 10 Gyr), while BS stars are likely similarly old, but less age-constrained. Therefore, there is a possibility of younger stars being included in this photometric selection, but the exact age distribution is difficult to determine. They also mask two regions of the sky to reduce the influence of the Virgo Overdensity and Sagittarius stream on the recovered parameters \citep{Bell2008}. With these stars, they find the halo between 7 kpc$<r<$145 kpc is best described by a smooth, oblate broken power law with an inner slope of $\alpha_\mathrm{in} \sim -2.3$, an outer slope of $\alpha_\mathrm{out} \sim -4.6$, a break radius at $\sim 27$ kpc, and a flattening of $q \sim 0.6$. 

\subsubsection{\citet{Iorio2018}}
\citet{Iorio2018} use $\approx$21,600 RR Lyrae (RRL) stars selected photometrically from a cross-match between Gaia Data Release 1 and 2MASS to study the morphology of the MW stellar halo. RRLs are old ($\gtrsim10$ Gyr), metal-poor pulsating stars with negligible contamination from younger populations, making them a clean age-constrained tracer of the ancient halo \citep{Smith1995}. They remove two regions of the sky with radius 7$^{\circ}$ and 9$^{\circ}$ for the SMC and LMC, respectively. Using maximum likelihood fitting, they find the inner halo density profile is well described by a single power law with $\alpha = -2.96 \pm 0.05$, with no significant evidence for a break within their radial range ($r=1.5-30$ kpc). 

\subsubsection{\citet{Lucey2026}}
\citet{Lucey2026} also use RRL, but expand the sample significantly to $\sim105,000$ stars drawn from Gaia DR3, ASAS-SN, and PanSTARRS1, spanning galactocentric radii of $\approx 0.2-120$ kpc. Similar to \citet{Iorio2018}, the region of the sky with the SMC and LMC is removed from the sample. Rather than a parametric power-law model, they employ a hierarchical Bayesian Gaussian Mixture Model, which provides the flexibility needed to simultaneously model the overlapping bulge, disk, and halo components in the inner Galaxy. They find the outer halo ($r \gtrsim 10$ kpc) follows a $r^{-4}$ power law with flattening $q = 0.70$ and a tilt of $18^\circ$. The inner Galaxy ($r \lesssim 10$ kpc) is dominated by a distinct prolate population with $q = 1.31$, inconsistent with a simple extrapolation of the outer halo profile. 

\subsection{Chemically Selected Samples}

The following works use a variety of methods to select their halo sample. \citet{Han2022} specifically targets accreted stars from the GES system, while \citet{Kurbatov2024} employs a simple metallicity selection. On the other hand, \citet{Yang2022} aims to merely remove the disk stars from their sample of K giants with a cut in rotational velocity ($v_{\phi}$) and metallicity. According to \citet[][see Figure 19]{Naidu2020}, the relative contribution of GES to the stellar halo peaks at galactocentric radii $\sim$20 kpc and decreases towards small radii. Therefore, it is likely that \citet{Han2022} will be biased toward measuring a shallower profile at these radii relative to the total halo stellar mass profile. Similarly, the \citet{Kurbatov2024} profile will be biased towards shallower slopes if the more metal-rich population is more centrally concentrated. Given the complicated age-metallicity relation for halo stars \citep{Bonifacio2026}, this is related to but distinct from the bias from age-constrained tracers. By performing orbit integration and minimal cuts to the sample selection, \citet{Yang2022} achieves the most direct measurement of the total stellar halo mass profile. It is possible there are still biases that may impact this work, (e.g., if much of the local halo has similar velocities and metallicities to the disk and the assumption of the Milky Way potential model likely has non-negligible effects), but they are not easily quantified without a detailed mock study and are certainly distinct from the biases of the other works to which we compare. Below, we discuss each of these works in more detail.    

\subsubsection{\citet{Han2022}}

\citet{Han2022} take a different approach, using a chemically selected sample rather than an age-constrained tracer. Specifically, they use chemical abundances  and kinematics from the H3 Survey to identify 5,559 giant stars belonging to the GES merger in the radial range $r = 6-60$ kpc, selecting stars with orbital eccentricity $e > 0.7$, $[\alpha/\mathrm{Fe}] < 0.27 - 0.5([\mathrm{Fe/H}] + 0.7)$, and $[\alpha/\mathrm{Fe}] < 0.27 - 0.1([\mathrm{Fe/H}] + 0.7)$. For comparison to other works, this effectively restricts the sample to [Fe/H]$\leq$-0.5. We also note they perform a cut on the angular momentum to remove the impact of the Sagittarius stream. Forward modeling the full H3 selection function, they fit a general triaxial ellipsoid with principal axes allowed to rotate freely with respect to the Galactocentric frame, coupled with a multiply-broken power law. They find the halo is best described by a near-prolate triaxial ellipsoid (axes ratios 10:8:7) tilted $25^\circ$ above the Galactic plane, with a doubly-broken power law with power law slopes of $\alpha=-1.7$, $-3.1$, and $-4.6$  and break radii at $\sim 12$ kpc and $\sim 28$ kpc. We note this work also validates their methodology on mock data. 

\subsubsection{\citet{Yang2022}}

\citet{Yang2022} take a novel approach, using orbit integration rather than a direct density measurement to probe the halo profile down to the Galactic center. They select $\sim$10,000 halo K giants at 5 kpc $\leq r \leq 50$ kpc from LAMOST DR5 cross-matched with Gaia DR3. Halo membership probabilities are defined by fitting a two-component skewed normal distribution to the joint [Fe/H]--$v_{\phi}$ plane, where one component traces the disk and the other the halo, effectively excluding stars with $-300\ \mathrm{km/s} \lesssim v_{\phi} \lesssim -30\ \mathrm{km/s}$ and $ -1.5 \lesssim \mathrm{[Fe/H]} \lesssim 0.5$ \citep[see Figure 1 in][]{Yang2022}.  Stars with a halo membership probability greater than 0.6 are retained as halo members. Importantly, stars of all metallicities are included provided their $v_\phi$ is inconsistent with the disk distribution. With this sample in hand, they integrate stellar orbits in a MW potential to construct the full density distribution down to $r \lesssim 1$ kpc. They find the halo is well described by a double-broken power law with slopes of $\alpha = -1.5$, $-2.8$, and $-6.1$ at $r<10$ kpc, 10 kpc $< r < 25$ kpc, and $r>25$ kpc, respectively.

\subsubsection{\citet{Kurbatov2024}}

\citet{Kurbatov2024} focuses on the inner MW, using red giant branch (RGB) stars from Gaia DR3 with metallicities estimated from XP spectrophotometry to model the spatial density distribution of metal-poor ([M/H] $< -1.3$) stars, spanning Galactocentric distances from $\sim 1$ to 18 kpc. Carefully accounting for Gaia's selection function, they fit a single-component cored power-law model, finding a vertical flattening of $q\approx 0.5$ and a power-law slope of $\approx -3.4$, consistent with previous studies \citep{Pietrukowicz2015,Iorio2018}. They also fit a two-component model to allow for greater flexibility; however, we do not compare to this model as the relative contributions of the two components as a function of Galactocentric radius are not publicly available. 

\section{Milky Way's Broken Profile Measurements in Tension with Cosmological Simulations} \label{sec:break}

\begin{figure}
    \centering
    \includegraphics[width=\linewidth]{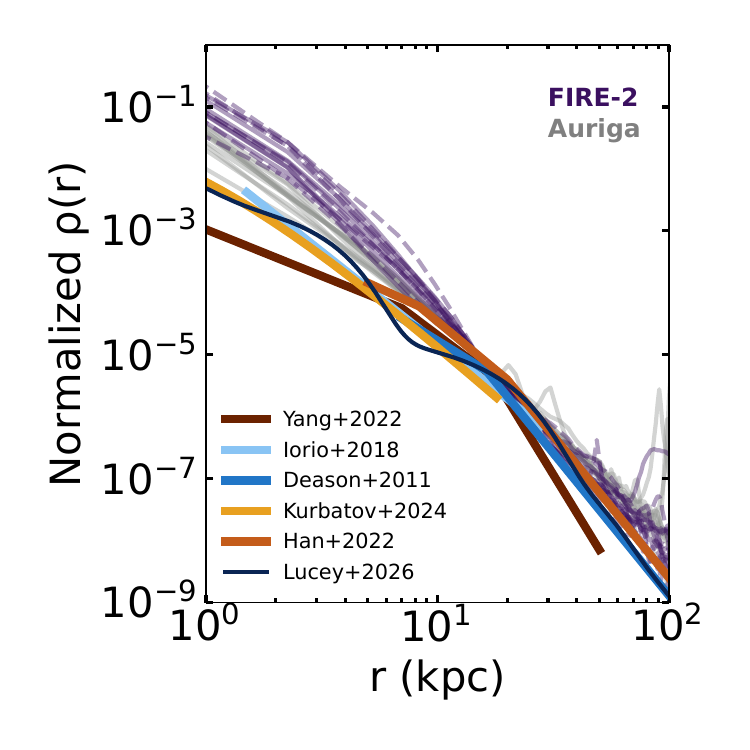}
    \caption{The halo profiles of galaxies from two cosmological zoom-in suites, FIRE-2 (purple) and Auriga (grey) compared to measurements of the MW's halo profile. The simulations profiles are calculated by selecting particles within a conical volume centered on the $z$-axis defined  to be perpendicular to the galactic disk. The six FIRE-2 galaxies from the \textit{ELVIS} suite of Local Group-like pairs are shown as dashed lines. The MW literature results shown and discussed in this work include \citet[][brown]{Yang2022}, \citet[][light blue]{Iorio2018}, \citet[][blue]{Deason2011}, \citet[][light orange]{Kurbatov2024}, \citet[][orange]{Han2022}, and \citet[][dark blue]{Lucey2026}. To fairly compare to the simulated profiles, we specifically look at the steepest part of the measured 3D radial profiles which is along the Galactic $z$-axis where $r=\sqrt{(z/q)^2}$. All profiles are normalized to be approximately equal at r=15 kpc, with small offsets for visibility.  The MW measured profiles become shallower around $z=$10-20 kpc, diverging from the simulated galaxy halo profiles.   }
    \label{fig:profs}
\end{figure}

Early numerical work has shown that halo profile breaks similar to the MW can be caused by apocenter pile-up of accreted systems \citep{Deason2013}. In this section, we investigate whether this result holds in cosmological zoom-in simulations, specifically the FIRE-2 and Auriga MW-mass suites. 

In Figure \ref{fig:profs}, we directly compare the halo profiles of the cosmological simulations to results from the MW literature, which are described in detail in Section \ref{sec:MWLit}. Here and throughout this work, simulated halo profiles are computed using particles within a double cone of half-opening angle $\pi/4$ centered on the galactocentric minor ($z-$) axis (i.e., $\theta \leq \pi/4$, where $\theta$ is the polar angle from the $z-$axis), in order to exclude disk particles without introducing biases from metallicity or kinematic selections. In order to fairly compare the MW measurements to this given the flattening ($q$), we specifically look at the $z-$axis of the halo profiles in that $r=\sqrt{(z/q)^2}$.  As we are most interested in comparing the slopes of the profiles rather than the absolute stellar density, which is more difficult to measure observationally, each simulated profile is normalized be equal at r=15 kpc. The observational profiles are normalized to be approximately equal at r=15 kpc, with small offsets introduced so that each line remains visible while the profiles still agree well at r>15 kpc. This radius was chosen somewhat arbitrarily, based on its central position in the figure and on where that agreement holds.  The FIRE-2 results are shown in purple with the $\textit{ELVIS}$ Local Group-like pairs shown with dashed lines and the \textit{Latte} suite with solid lines. The Auriga results are shown in grey. Throughout this work, the age-constrained MW literature results are shown in shades of blue, with \citet{Lucey2026} in dark blue, \citet{Deason2011} in blue, and \citet{Iorio2018} in light blue. The chemically selected results are shown in shades of brown/orange, with \citet{Yang2022} in brown, \citet{Han2022} in orange, and \citet{Kurbatov2024} in light orange. 

We do not remove any satellites when calculating the profiles, which cause the large bumps seen in some profiles at r$\gtrsim$ 20 kpc. Although some of the literature observational works do remove substructures to reduce a bias in their global halo fit, since we do not fit the profiles and instead just plot the stellar mass, removing substructure would only impact the profiles if they contribute a significant fraction of the stellar mass at that radii. This would  remove some of the bumps at large radii which are likely intact satellites, and would be interesting future work to investigate if massive stellar streams similar to Sagittarius impact these profiles.   We note that both the simulated and measured MW profiles are agnostic to the difference between bulge and halo. The simulated profiles are simply the stellar density profile along the minor-axis. The measured MW profiles that reach into the inner galaxy (r$\lesssim 5$ kpc), \citet{Yang2022,Iorio2018,Kurbatov2024,Lucey2026}, are also agnostic to the separation of bulge and halo, and model the stellar density profile as one continuous component, which is consistent with the simulations.

The cosmological simulation halo profiles do not show a break and are much steeper than the MW profiles within $r<15$ kpc. The FIRE-2 simulations generally show slightly steeper profiles in this region compared to the Auriga profiles, but both are considerably steeper than the measured MW  profiles. The results of \citet{Lucey2026}, \citet{Iorio2018} and \citet{Kurbatov2024} are closest to the simulated profiles in the inner region, but are still significantly shallower. \citet{Iorio2018} and \citet{Kurbatov2024} do not find a break and focus only on the inner halo, with maximum galactocentric radii of 30 kpc and 18 kpc, respectively.  \citet{Lucey2026}, on the other hand, finds a shallower slope consistent with the breaks of \citet{Han2022}, \citet{Yang2022} and \citet{Deason2011}, but finds that the profile steepens again inside $r<10$ kpc, which is also inconsistent with the cosmological simulations.  

It is important to note that throughout this work we compare normalized stellar density profiles, meaning we are sensitive to the slopes of the profiles rather than the absolute number of stars. Using the same simulations, \citet{Lucey2025} found that the inner halo halo density slope is strongly correlated with the dark matter profile slope, rather than the merger history of the host galaxy. Specifically, the ratio of stellar mass density to dark matter density falls off as 1/$\rm{r^2}$ in each galaxy. Therefore, it is possible that the discrepancy between the inner stellar halo profile slopes of the simulated and Milky Way profiles is related to dark matter physics rather than galaxy formation processes. 

\subsection{Simulated Halos Rarely Break, and Never as Sharply} \label{sec:nobreak}

\begin{figure*}
    \centering
    \includegraphics[width=\linewidth]{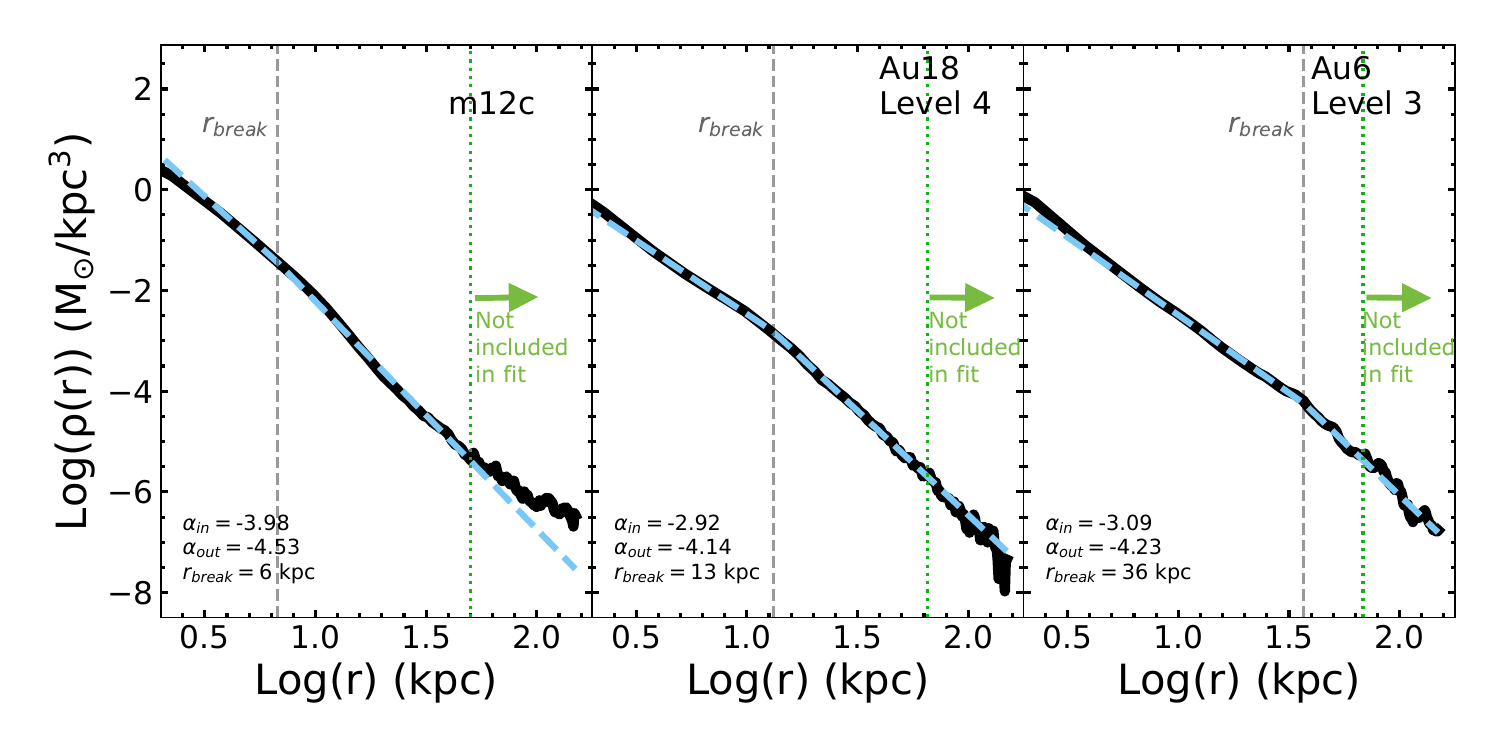}
    \caption{The halo profiles of the three simulated galaxies with the strongest breaks, m12c from the FIRE-2 suite and the Auriga galaxies 18 (resolution level 4) and 6 (resolution level 3) . The black line is the stellar halo profile while the light blue dashed line is the broken power-law fit. The fit parameters, inner slope ($\alpha_{in}$), outer slope ($\alpha_{out}$) and break radius ($r_{break}$), are printed in the lower left corners of each panel. The grey dashed line shows the break radius and the green dotted line shows the maximum radius used in the fit from the rising mask.  While these galaxies statistically favor the broken profile fit over a single power-law, the breaks are not as strong as the MW (i.e., the difference between the inner and outer slopes is smaller).   }
    \label{fig:broke_profs}
\end{figure*}

To further investigate the prevalence of breaks, we fit single and broken power law profiles to the stellar halos of MW-mass cosmological zoom-in simulations. In order to minimize the impact of substructure and Poisson noise in the outer halo where there are small numbers of star particles, we utilize a rising mask, where if the density increases by more than 10\% in the  neighboring larger radial bin ($\Delta$ r =1 kpc), we do not include data beyond that radius in the fits. The rising mask also serves to focus the fitting power on the inner galaxy (r$\lesssim50$ kpc), as the intent is to determine if they have breaks similar to the MW's which occurs at r$\approx 20-30$ kpc. This mask results in an average maximum radius of r$\approx75$ kpc for the fitting procedure.    

To determine whether a broken power law provides a statistically significant improvement over a single power law, we employ an F-test on the residual sum of squares, which is well-suited for comparing nested models in the absence of formal measurement uncertainties. The break location $r_\mathrm{break}$ is only defined under the more complex model, which biases the standard F-test toward overstating significance (yielding p-values that are systematically too small). We therefore adopt a stricter threshold of $p < 0.01$, rather than the conventional $p < 0.05$, to compensate. There are five total galaxies that satisfy this criteria and have statistically significant breaks between 5 kpc $<r<$ 70 kpc. We focus on three of these galaxies where the differences between the outer and inner power-law slopes are $>$ 0.5.

The Auriga stellar halo profiles have also been studied in \citet{Monachesi2019}. From visual inspection, \citet{Monachesi2019} estimates that 20\% of the Auriga profiles have breaks, which is consistent with our findings that 2 out of the 11 (18\%) Auriga galaxies we study have significant breaks. When we include FIRE-2, the percentage drops to 12.5\% with 3 out of 24 simulated halo profiles having breaks. 

In Figure \ref{fig:broke_profs}, we show the broken profile fits for the three galaxies with statistically significant strong breaks, m12c from the FIRE-2 suite and Auriga galaxies 18 and 6, run at resolution levels 4 and 3, respectively. The normalized density profiles are shown as a dark black line while the fit profiles are shown as a light blue dashed line. The break radius ($r_{break}$) is shown as a vertical grey dashed line and is printed in the lower left corner along with the inner and outer slopes ($\alpha_{in}$ and $\alpha_{out}$). The green dotted line indicates the fit’s outer radial limit, defined by the rising mask where the density increases by more than 10\% from one radial bin to the next. 

With inner halo slope estimates between $\alpha_{in}$ $\approx -1.5$ and $-2.9$ and outer halo slope estimates between $\alpha_{out}$ $\approx -4.5 $ and $-6.1$, the MW halo break is reported to be strong, where the difference between the inner and outer slope is  $\sim$1.6-4.6 \citep{Deason2011,Han2022,Yang2022,Medina2024,Amarante2024}. None of the breaks fit to the cosmological simulations are nearly as strong, with the largest difference seen in the Auriga 18 run at resolution level 4, which has a slope difference of only 1.22. This provides further evidence for a tension between the cosmological simulations and MW observations.

\subsection{The Breaks are not Predominately Apocenter Pile-up} \label{sec:gse}

\begin{figure*}
    \centering
    \includegraphics[width=\linewidth]{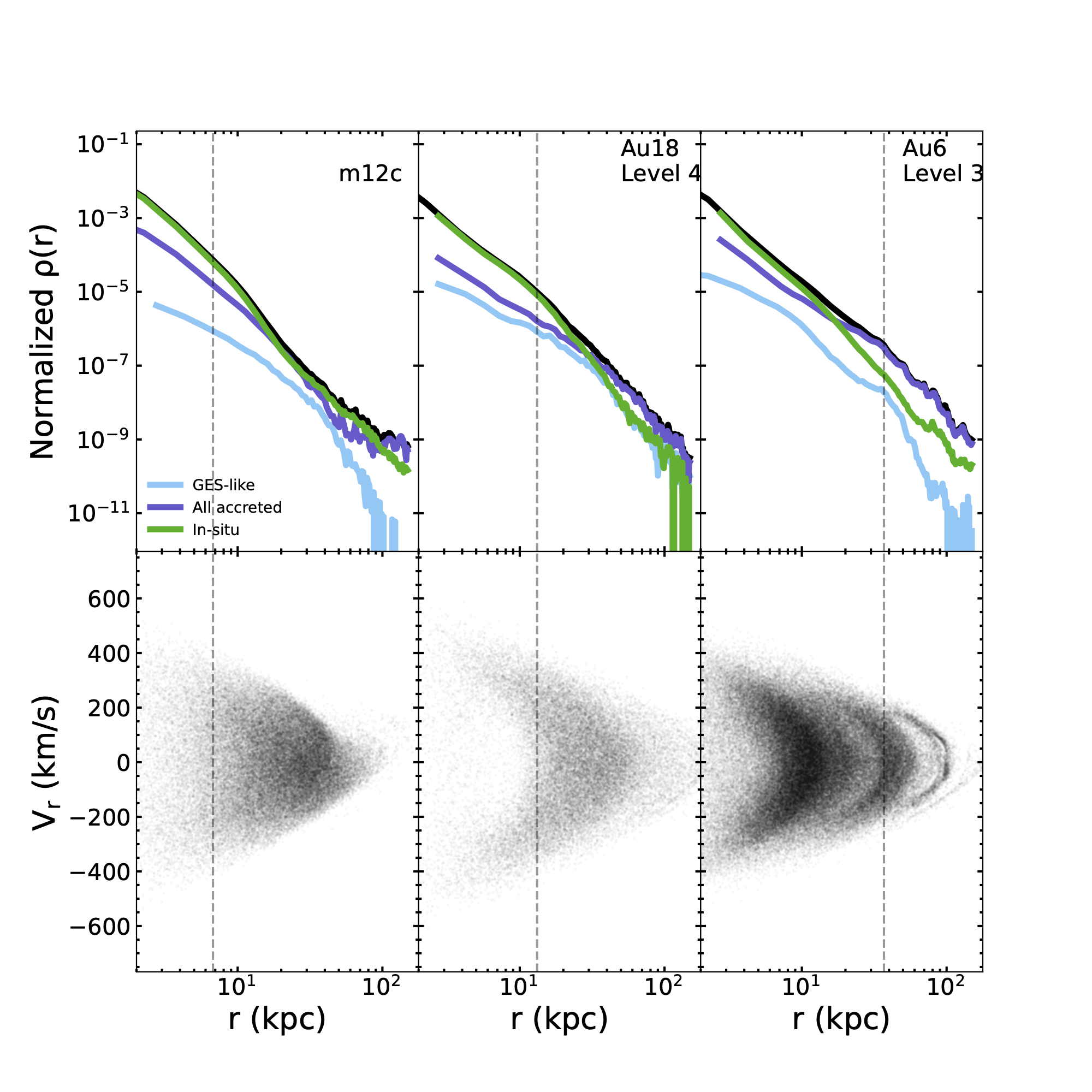}
    \caption{The top panel shows the stellar halo profiles of the three simulated galaxies with strong breaks, broken into components based on their origin. The black line shows the entire halo profiles, while the profile calculated using only stars that formed in-situ is shown in green and for stars that were accreted is shown in purple. The light blue line shows the halo profile for a specific accreted galaxy that is chosen to be GES-like. The vertical grey dashed line shows the break radius. Auriga 6 is the only galaxy whose break occurs at a radius where the accreted stars dominate the halo profile. In the bottom panels, we show the radial velocities as a function of galactocentric radius for the stellar particles associated with the accreted GES-like system. For Auriga 6, the break radius aligns with the apocenter pile-up of this merger remnant.  }
    \label{fig:gse_profs}
\end{figure*}

The break in the MW's halo density profile is thought to be caused by apocenter pile-up of the accreted GES stars  \citep{Deason2011,Deason2018}. To investigate whether we see this effect in cosmological hydrodynamic zoom-in simulations, we identify GES-like systems in the three simulated galaxies with the strongest profile breaks. To do this, we first check available literature which has calculated the radial anisotropy of merger remnants and used that to identify GES-like systems. M12c does not have a merger remnant with radial anisotropy >0.65. Instead, we investigate the merger with a mass ratio and timing similar to estimates of GES.  The merger we identify as GES-like in m12c has a mass ratio of 0.35 and redshift of 1.34, while GES is estimated to have been a mass ratio of 0.24 and occurred at a redshift of $z\approx 1.8$ \citep{Helmi2018}. For Auriga 18 run at resolution level 4, we use a GES-like merger identified in \citet{Fattahi2019} based on radial anisotropy and contribution to the stellar halo. Auriga 6 run at level 3 is not studied in \citet{Fattahi2019}, which only investigates the level 4 simulations. The Auriga 6 level 4 galaxy does have a radial merger but it is not identified as a GES analog in \citet{Fattahi2019}, because it does not contribute a large fraction of the stellar halo. Instead, we investigate the density profiles of four of the most massive mergers that occurred at redshifts $\lesssim$ 2. We chose the GES analog as the system with the strongest break in the density profile. 

In the top panels of Figure \ref{fig:gse_profs}, we show the GES analog density profiles (light blue), along with the entire stellar density profile (black), the profile of the accreted stars (purple) and the in-situ stars (green). In the bottom panels, we show the galactocentric radial velocities ($v_r$) as a function galactocentric radius ($r$). For all the panels, the measured break radius in the overall stellar density profile is shown as a grey dashed line. For m12c and Auriga 18 at level 4, the total stellar profile breaks occur at radii dominated by the in-situ stellar population. For these galaxies, the total stellar profile breaks are related to a steepening in the in-situ stellar population profile, similar to the halo breaks in \citet{Font2020}. The $v_r$-$r$ plane demonstrates that the majority of apocenters of the GES-like systems for these galaxies are also significantly larger than the total stellar profile break radius. On the other hand, Auriga 6 at level 3 has a total stellar profile break at a radius that is dominated by accreted stars. We also identified an accreted system that has a strong break in its density profile at that radius. Based on the $v_r$-$r$ plane, this break is caused by apocenter pile-up. 

Breaks in the total stellar halo profile are rarely caused by apocenter pile-up in cosmological hydrodynamic simulations, appearing in only 1 of 21 unique simulated halos (three Auriga halos are run at both resolutions, giving 24 total profiles). In-situ material dominates the stellar density profiles at $r\lesssim 20$ kpc, meaning any break in the total stellar density profile that is due to apocenter pile-up must occur at larger radii. This is consistent with the single galaxy in our sample that exhibits such a break, which occurs at $r=36$ kpc. We note that the profiles of the individual GES-like systems all show breaks related to apocenter pile-up, consistent with \citet{Amorisco2017}, but the dominance of the in-situ population causes these breaks to not impact the total stellar halo profile.

\section{Milky Way Consistent with Accreted Profiles} \label{sec:accr}

\begin{figure*}
    \centering
    \includegraphics[width=\linewidth]{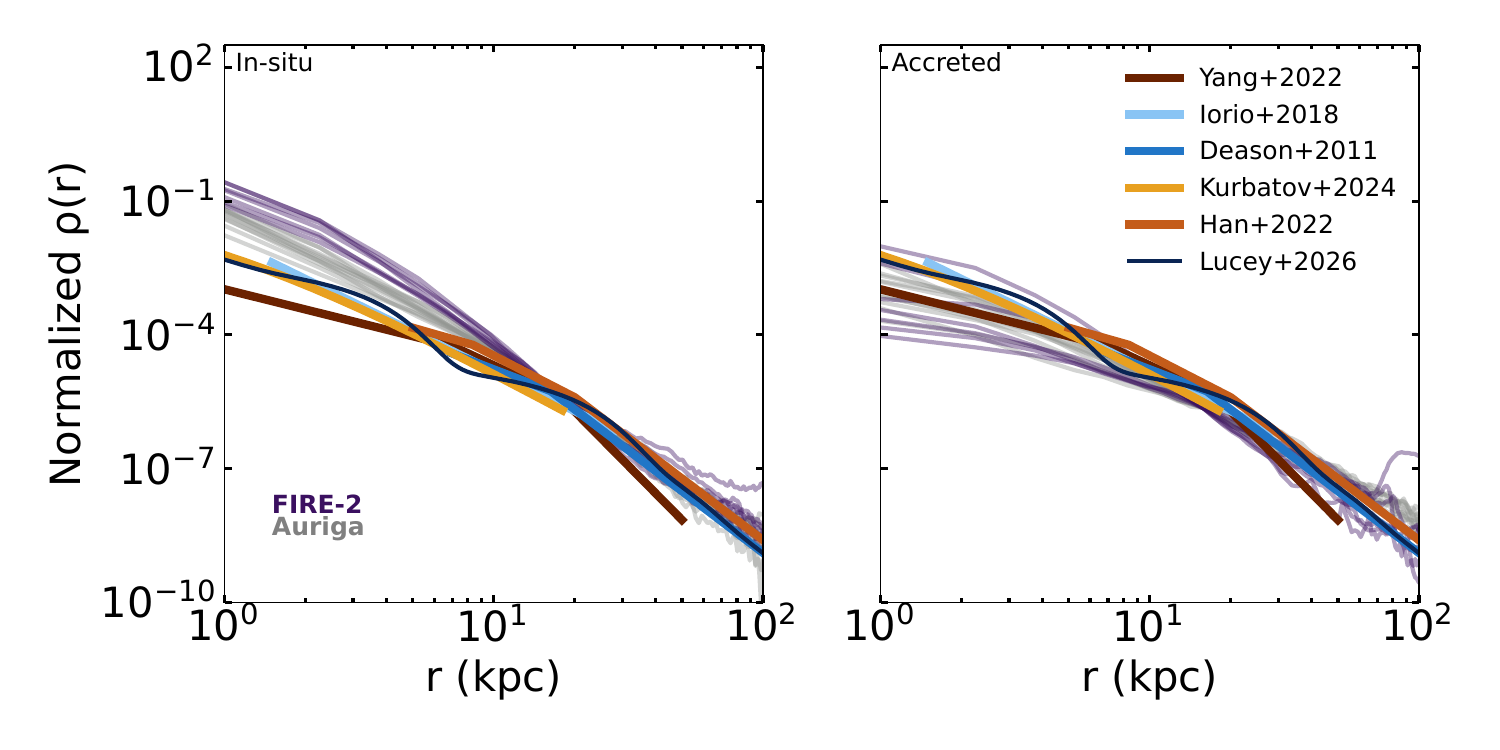}
    \caption{The simulated galaxy profiles and literature results are the same as in Figure \ref{fig:profs}, except the simulated profiles are divided up by origin, with in-situ stars in the left panel and accreted stars in the right panel. The accreted stellar profiles of the simulations are consistent with the MW measurements, while the in-situ stellar profiles are steeper than the MW measurements within $\approx$ 15 kpc of the Galactic center.   }
    \label{fig:accr_profs}
\end{figure*}

\begin{figure}
    \centering
    \includegraphics[width=\linewidth]{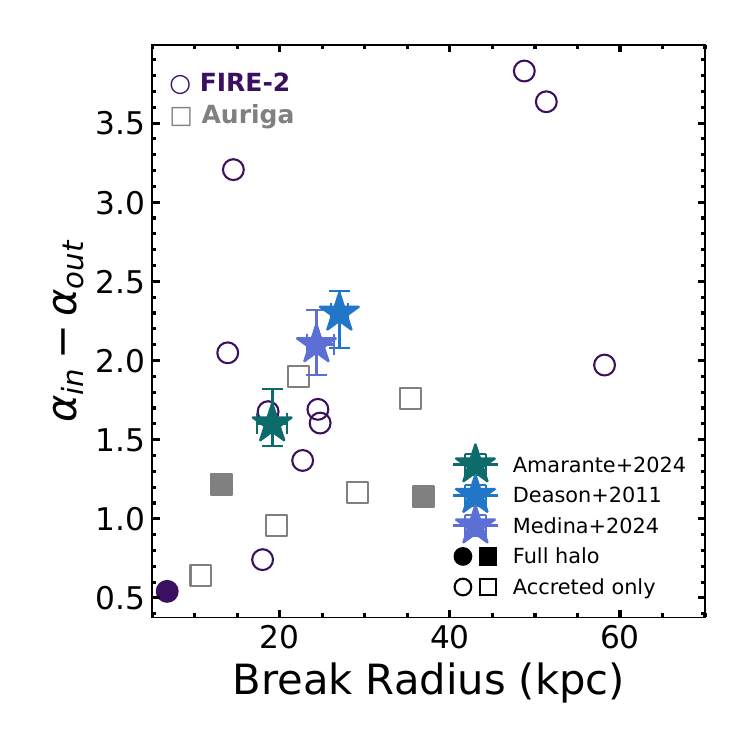}
    \caption{Estimates of the MW's break strength ($\alpha_{in}-\alpha_{out}$ as a function of the break radius, compared to FIRE-2 (purple) and Auriga (grey) simulations. The MW results are from works that find double power-law fits with \citet{Amarante2024} in blue-green, \citet{Deason2011} in light blue and \citet{Medina2024} in light purple. The three simulated galaxies with statistically significant break in their stellar halo profiles are shown as filled circles, while the same for only the accreted star profiles are shown as open circles. The simulated galaxy breaks are weaker than the lowest estimates for the MW's break, but the accreted stellar profiles show breaks consistent with MW estimates.  }
    \label{fig:breaks}
\end{figure}

In Figure \ref{fig:accr_profs}, we isolate the accreted and in-situ contributions to the profiles of the cosmological simulations. Specifically, we show the in-situ halo profiles for the simulations in the left panel, while the right panel shows the profiles for the accreted stars. The literature results and colors are the same as in Figure \ref{fig:profs}. It is interesting to note that in the simulated galaxies, the in-situ population contributes significantly to the stellar halo even at large radii (r$\approx$ 100 kpc). It is possible that this is at least partially driven by imperfect classification of accreted stars, because of the limits of numerical (sub)halo tracking. Furthermore, the edge case of stripped accreted gas that forms stars is not classified as accreted. It is also possible that these in-situ stars formed early in the outer halo. Given the difficulty to observationally distinguish in-situ and accreted populations, it is unclear if in-situ populations significantly contribute to the Milky Way's profile at these radii, but this could perhaps be investigated in future work by comparing the stellar mass in the smooth distribution to substructure.   

The accreted stellar halo profiles of the simulations align well with the measured MW profiles.  To explore this agreement further, we also fit the accreted profiles with single and broken power laws. Unlike for the total (in-situ + accreted) profiles, almost all of the FIRE-2 galaxies statistically prefer the broken profile fit, with the exception of m12m, Louise and Juliet. M12m is an outlier in many respects (see Section \ref{sec:origins}), including not having a major (>5\% mass ratio) merger since before its disk formed. Juliet and Louise, on the other hand, is exceptional because of major substructure in the accreted profile, which impacts the fitting procedure. For the Auriga galaxies accreted stellar profiles, we find that 6 out of the 11 galaxies do not have statistically strong breaks. However, of these 6, two are Auriga 24 and two are Auriga 27, run at different resolution levels. Auriga 23 is also included at both resolution levels, but only Auriga level 4 does not have a break. The last Auriga galaxy without a break in the accreted profile is Auriga 21 level 3. Of the five Auriga galaxies with statistically significant breaks, two (Auriga 18 level 4 and Auriga 6 level 3) have been previously discussed as having significant breaks in their total stellar profiles. Furthermore, Auriga 10 level 4 has a strong break and is also identified in \citet{Fattahi2019} as having a GES-like merger. Auriga 16 level 3 also has a strong break, but is not studied in \citet{Fattahi2019}.

Figure \ref{fig:breaks} shows the estimates of the strength of the MW's halo break as quantified by the difference between the inner and outer slopes ($\alpha_{in}-\alpha_{out}$), compared to the Auriga and FIRE-2 MW-mass galaxies. We specifically show literature values from three works which found the double power-law as the best fit, \citet{Amarante2024} in teal, \citet{Deason2011} in blue and \citet{Medina2024} in periwinkle. We do not include results for works which found triple power-laws on this plot since it is not a 1-to-1 comparison. We note the difference between the innermost and outermost slopes for those works range from 2.9-4.6 \citep{Han2022,Yang2022,Li2026}. For the Auriga and FIRE-2 simulations, we show break strengths and radii for the three whose stellar halo had statistically significant breaks as discussed in Section \ref{sec:nobreak}. These are shown as grey (Auriga) and purple (FIRE-2) filled circles. The break qualities for only the accreted halos are shown in open circles for the 10 FIRE-2 galaxies and 5 Auriga galaxies with statistically strong breaks in their accreted profiles. 

When the in-situ stars are included, the simulations' halo profile breaks are all weaker than even the lowest estimates of the MW's break. The accreted profiles, however, show a large range of break strengths and radii, including stronger breaks than MW estimates. The accreted profile breaks with strengths $>2$ are all from FIRE-2 simulations. Given that the Auriga accreted profiles also have a lower fraction of statistically significant breaks, we find that FIRE-2 simulations produce systematically stronger breaks in their accreted halo profiles than Auriga simulations.

Taken together, these results suggest it is entirely the in-situ stars in the simulated halos that are discrepant with MW measurements; these stars dominate the halos within galactocentric radii of 10-20 kpc. This is consistent with \citet{Monachesi2019}, who also find the Auriga simulations at resolution level 4 better match the stellar halo profiles of nearby galaxies when excluding in-situ stars. We note that the recent FOGGIE simulations have more concentrated in-situ stellar halo profiles, and it would be interesting to see if they match the MW profile better \citep{Wright2024}. \citet{Han2022} intentionally target the accreted halo, but all of the other literature results do not, although they have their own selection effects which are discussed further in Section \ref{sec:biases}. This result compels us to wonder: where is the MW's in-situ halo? Or are the simulations overproducing an in-situ halo component?

As discussed in Section \ref{sec:break}, it is possible that the discrepancy identified here reflects dark matter physics rather than galaxy formation processes. In that case, the apparent connection between the in-situ stellar population and the profile discrepancy may be a coincidence. It is possible that the radius within which the in-situ stars dominate the simulated stellar halo profiles just happens to align with the MW's profile breaks, rather than the in-situ component being the true physical driver. The underlying cause of the discrepancy may instead be a more cored dark matter profile in the MW relative to the simulations. We discuss additional possibilities in the following sections.

\subsection{Impact of Sample Selection Biases on Measured Profiles} \label{sec:biases}

\begin{figure*}
    \centering
    \includegraphics[width=\linewidth]{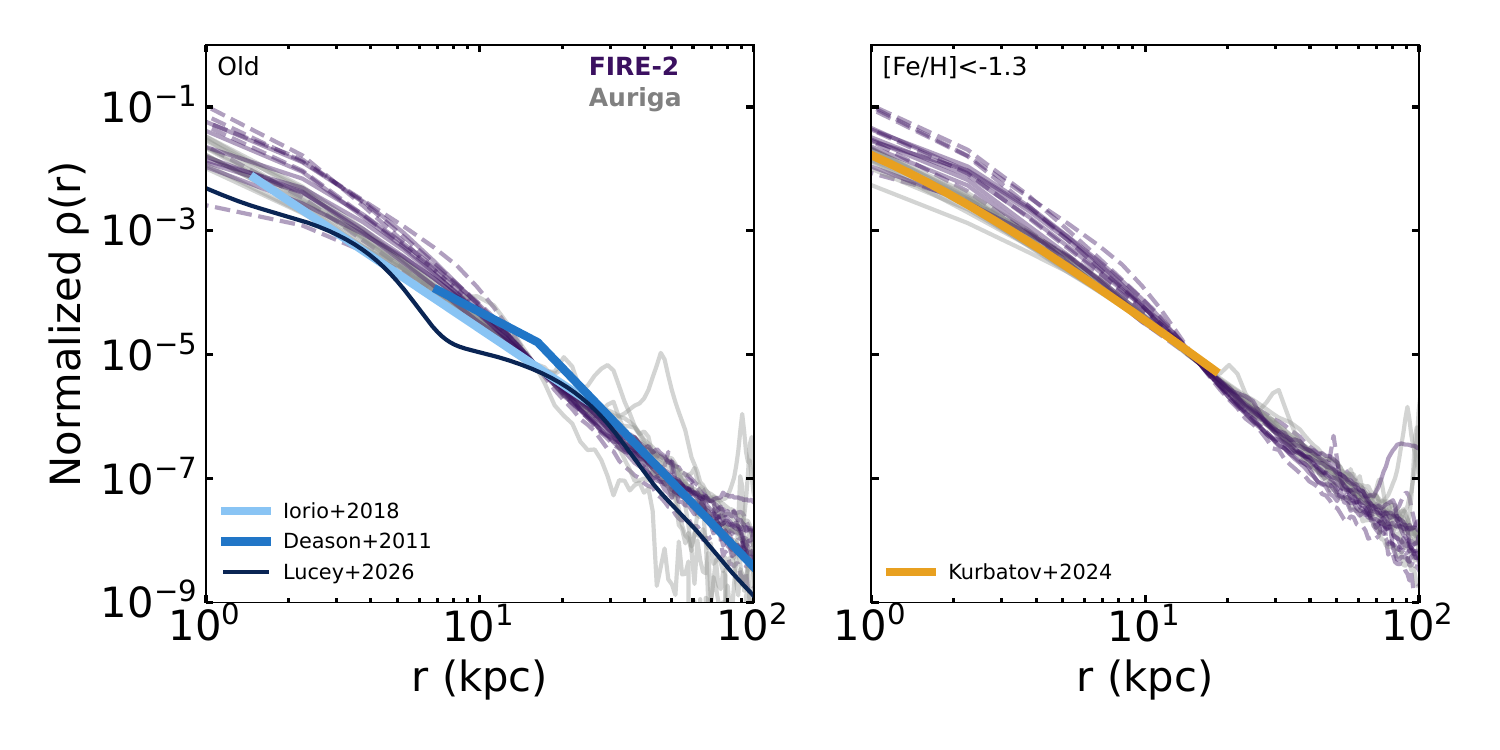}
    \caption{ The impacts of different selections on the halo profile. In the left panel, we directly compare the age-constrained literature results \citep{Iorio2018,Deason2011,Lucey2026} to the simulated profiles for stellar particles with the corresponding age-constraint ($>10$ Gyr). There is better agreement when matching the selection of these literature results but the difference is not fully resolved. The right panel specifically investigates the impact of the metallicity selection of \citet{Kurbatov2024}. When we restrict the simulated halo profiles to only stars with [Fe/H]<-1.3, we find better agreement with the \citet{Kurbatov2024} result, although the FIRE-2 profiles are generally still a bit steeper.   }
    \label{fig:profs_age}
\end{figure*}

As discussed in Section \ref{sec:MWLit}, each of the MW measurements use different stellar tracers and methods that can impact the results. In general, the literature can be divided into two categories: age-constrained standard candles and chemically-selected samples. Of the wide variety of works we discuss in this work, the three age-constrained results all primarily trace stars with ages $>10$ Gyr. \citet{Lucey2026} and \citet{Iorio2018} use RR Lyrae, while \citet{Deason2011} use BHB and BS stars, which are less age-constrained than RR Lyrae but are still primarily older than 10 Gyr. 

In the left panel of Figure \ref{fig:profs_age}, we compare these literature results with the halo profiles of stellar particles with ages $>10$ Gyr in the cosmological simulations. While the discrepancy between the simulations and the MW profiles is lessened, it is not completely alleviated. The simulated old stellar halo profiles still show an overabundance in the inner $\approx$20 kpc relative to the measured MW profiles. 

In the right panel of Figure \ref{fig:profs_age}, we also explore the impact of selecting metal-poor stars on the derived halo profile. Here, we specifically, recreate the metallicity selection of \citet{Kurbatov2024}, who measure the profiles of stars with [Fe/H] <-1.3. Consistent with previous works \citep{Fattahi2019,Carrillo2024}, we have applied a $-0.5$ dex offset to the Auriga metallicities which are known to be too metal-rich relative to the Milky Way. With this correction, we find the results are consistent with \citet{Kurbatov2024}, especially the Auriga simulations, while the FIRE-2 simulations still show a small overabundance within $\approx$ 10 kpc of the Galactic center. Therefore, there is not a strong discrepancy between the simulations and the MW's metal-poor halo profile. The disagreement is primarily at higher metallicities. It is possible that this disagreement is at least partially driven by biases against metal-rich stars in the tracers and selections used for targeting the MW's stellar halo. Many works have intentionally tried to overcome this bias \citep[e.g.,][]{Yang2022}, but it is a difficult task. Further work with MW observations is required to completely rule out this possibility.  

\subsection{Origins of In-Situ Halo in Simulations} \label{sec:origins}
\begin{figure}
    \centering
    \includegraphics[width=\linewidth]{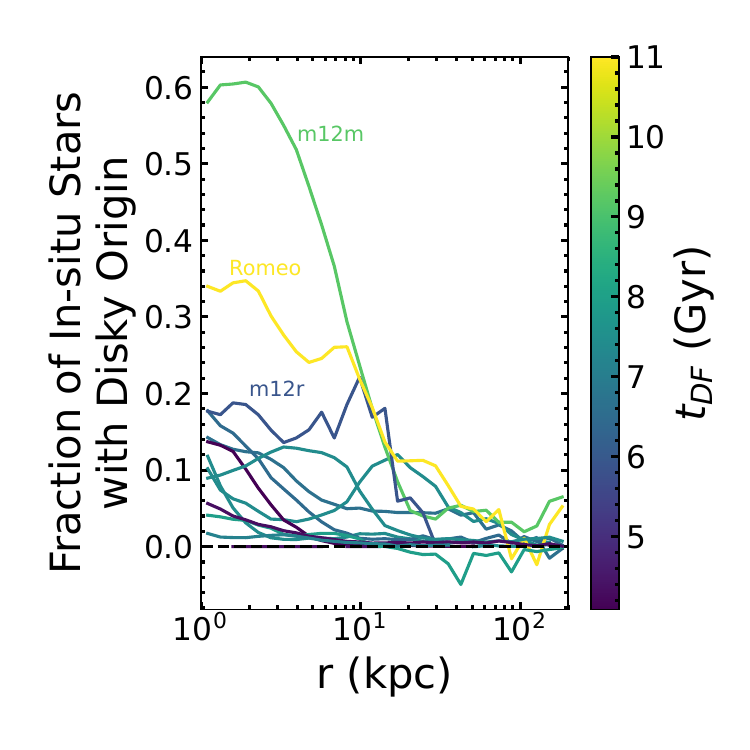}
    \caption{The fraction of in-situ halo stars that are born with disk-like kinematics as a function of galactocentric radius. The line for each galaxy is colored by its lookback time to the onset of disk formation, where larger lookback times have earlier forming disks. The largest outliers, m12m, Romeo and m12r, are labeled. The fraction of insitu stars with disky origin decreases with increasing galactic radii. For all except m12m, the majority of in-situ halo stars do not have disk origins at all radii.  }
    \label{fig:disk_frac}
\end{figure}

Generally there are thought to be two origins for in-situ halo stars: (1)  stars born through ELS-like dissipational contraction in the early universe where star formation occurs on halo-like orbits (2) stars that were born in the disk that are heated to halo orbits  \citep{Eggen1962, Zolotov2009,Font2011,Cooper2015,Samland2003, Purcell2010,Jean-Baptiste2017}.  To investigate the origins of the in-situ halo component that dominates the simulated galaxies at r$\lesssim$10--20 kpc, we calculate the fraction of in-situ stellar particles that were born with disk kinematics. For this part of the work, we focus only on the FIRE-2 simulations suites and exclude the Auriga galaxies because of the availability of published disk formation times for the FIRE-2 galaxies \citep{Yu2023,McCluskey2024}.  To select disk orbits, it is common to use the $z$-component of their angular momentum ($L_z$) where the $z$-direction is defined to have the highest net angular momentum  in order to capture the disk plane,  and $L_z>0$ is defined to be prograde and aligned with the disk rotation. However, prior to disk formation, without ordered rotation, the direction of highest net angular momentum can vary a lot with time \citep{Panithanpaisal2021}. Therefore, to investigate the disky origins of stars, we can really only make sense of their $L_z$ at formation ($L_{z,form}$)  if they form after the disk, when the $z$-direction can be more robustly defined. We instead estimate the number of  stars that are formed on disky orbits by first selecting in-situ stars that are younger than the lookback time corresponding to the onset of disk formation ($t_{DF}$), as calculated in \citet{McCluskey2024}, and that formed within a physical distance of 15 kpc from the galaxy's center. This is a conservative distance cut in that the disks are significantly smaller, even at z=0 \citep{Graf2025}.  Then, assuming the stars formed in the halo have a symmetric distribution in $L_{z,form}$, we subtract the number of stars with $L_{z,form}<0$ from the number of stars with $L_{z,form}>0$ and use the overabundance of $L_{z,form}>0$ as the number of stars born with disky kinematics. We divide this by the total number of in-situ halo stars within a given radius to get the disky origin fraction. We also compare this to the fraction of stars at a given radius with age <$t_{DF}$ and fraction of the total angular momentum that is aligned with the disk ($L_{z,form}/L_{form}$) >0.7. Both metrics give almost identical results.

In Figure \ref{fig:disk_frac}, we show how this disky origin fraction varies as a function of galactocentric radius for all of the FIRE-2 galaxies used in this work. The lines are colored by their disk formation lookback time as calculated in \citet{McCluskey2024}. In general, the fractions decrease with galactocentric radius, as expected. Negative fractions occur when there are more stars with $L_{z,form}<0$ than $L_{z,form}>0$ at a given radius. For most galaxies the fraction of in-situ halo stars with disky origins is $<$20\% at all radii. M12m and Romeo are the largest outliers with significantly higher fractions ar r$\lesssim$10 kpc. M12r is also a high outlier with a peak fraction of  $\approx$20\% at r$\approx$ 10 kpc. M12m and Romeo are also the galaxies with the earliest onsets of disk formation. For m12m the disk onset has a lookback time of 9.21 Gyr, not long after its last major merger (mass ratio 1:4) at a lookback time of 9.66 Gyr. Romeo has a disk onset lookback time of 11.00 Gyr and has an exceptionally quiescent merger history with no mergers with mass ratio $>$1\%. M12r is not an outlier for its time of disk onset ($t_{DF}=$5.9 Gyr), but it has the most recent major merger with a mass ratio of 0.39 and lookback time of 0.70 Gyr. In general, we find that the majority of the in-situ stellar halo stars are born on spheroidal orbits and not heated up from the disk into halo orbits. This is consistent with the Aquarius SPH re-simulations of \citet{Cooper2015}, in which heated disk stars make up a minority of the in-situ halo, and with Auriga, in which kicked-up disk stars contribute less than 20\% of the local stellar halo \citep{Orkney2026}. However, \citet{Khoperskov2023} find that the in-situ stellar halos in the HESTIA simulations are primarily made of heated disk stars. Observational works estimate about half of stars in the local kinematically-defined halo have thick disk-like chemistry \citep{Bonaca2017,Di_Matteo2019,Belokurov2020}. These stars are typically interpreted as heated disk stars, but this assumes that stars born on halo-like orbits would be chemically distinct from the thick disk, which is not well-justified.  

\section{Summary and Conclusions} \label{sec:sum}
The stellar halo encodes the mass assembly history of the MW, making its structural properties a key test of galaxy formation models  \citep{Bullock2001,Bullock2005,Helmi2008,Pillepich2014,Monachesi2019}. Observational studies consistently recover a broken power law density profile, with a break radius of $\approx20-30$ kpc and a shallower inner component with slopes ranging from $\alpha\approx-1.5$ to $-3$ \citep{Deason2011,Iorio2018,Han2022,Yang2012,Lucey2025}, though constraints within 10 kpc remain sparse and uncertain. While early numerical work suggests this broken profile may arise from the apocenter pile-up of accreted stellar material \citep{Deason2013}, it remains unclear whether this result holds in state-of-the-art fully hydrodynamical cosmological zoom-in simulations.

In this work, we compare the MW's stellar halo density profile with those of MW-mass galaxies in the FIRE-2 and Auriga cosmological zoom-in simulation suites. Specifically, we compare the observed and simulated profiles along the galactic $z$-axis, defined to be perpendicular to the disk. We then investigate the frequency and physical origins of broken halo profiles across these simulations. Finally, we leverage halo tracking catalogs to decompose the relative contributions of in-situ and accreted stars to the overall profile shape. We find:

\begin{itemize}
    \item Estimates of the MW's minor-axis profile are much shallower within 20 kpc than FIRE-2 and Auriga profiles, with power law slope estimates ranging from $\approx -1.5$ to $-3$, while the simulated profiles have a power law slope of $\approx -4$ on average in the same range.
    \item FIRE-2 and Auriga profiles rarely have breaks, and none exhibit breaks as strong as the MW's, with smaller changes in slope across the break.
    \item We find only one out of the 24  simulated galaxies has a break in its halo profile that is related to apocenter pile-up of an accreted system.
    \item The halo profiles of the simulated galaxies are dominated by in-situ stars at galactic radii of $\lesssim$ 20 kpc.
    \item The measured MW profiles match the profiles of \textit{accreted} halo stars in the simulated galaxies.
    \item The in-situ halo stars drive the discrepancy between the MW measurements and the simulated halos. 
    \item Dynamical heating of stars from the disk does not account for the majority of in-situ halo stars and typically contributes $<$20\% at radii $>1$ kpc.   
\end{itemize}

These results indicate a significant discrepancy between cosmological hydrodynamic simulations and the MW. It remains possible that improved, less biased measurements of the MW's inner halo will resolve this discrepancy. However, if the MW measurements continue to strengthen the case for a shallower inner halo then we will need to reconsider our galaxy formation theory, possibly looking for cosmologies that give more quiescent early formation histories or, consistent with the results of \citet{Lucey2025}, different dark matter density profiles. Observations of nearby galaxy stellar halos with the Nancy Grace Roman Space Telescope \citep{Sanderson2026}, as well as the \textit{Euclid} and \textit{Arrakihs} missions \citep{Arrakihs,Euclid},  will also help illuminate whether this tension reflects a true failure of cosmological simulations or whether the MW is an outlier . By determining whether other MW-mass galaxies exhibit similarly broken or cored halo profiles, Roman will provide a statistical context that is currently impossible from the MW alone, distinguishing between a systematic bias in MW halo measurements, a genuine shortcoming of current galaxy formation models, or simply a rare but valid outcome of hierarchical assembly.

\section*{Acknowledgements}

We thank David W. Hogg and Andrew Wetzel for insightful discussions that inspired parts of this work. This material is based upon work supported by the National Science Foundation under Award No. 2303831.
RES gratefully acknowledges travel support from the Simons Foundation under award 1018462.
AHR was supported by a fellowship funded by the Wenner Gren Foundation.
JS acknowledges support from program JWST-AR-06278 by NASA through a grant from the Space Telescope Science Institute, which is operated by the Association of Universities for Research in Astronomy, Inc., under NASA contract NAS 5-03127.

This work used the DiRAC@Durham facility managed by the Institute for Computational Cosmology on behalf of the STFC DiRAC HPC Facility (\href{https://dirac.ac.uk/}{www.dirac.ac.uk}).
The equipment was funded by BEIS capital funding via STFC capital grants ST/K00042X/1, ST/P002293/1, ST/R002371/1 and ST/S002502/1, Durham University and STFC operations grant ST/R000832/1.
DiRAC is part of the National e-Infrastructure.

\section*{Data Availability}
We have used simulations from the FIRE-2\footnote{\url{https://flathub.flatironinstitute.org/fire}} and Auriga\footnote{\url{https://wwwmpa.mpa-garching.mpg.de/auriga/data.html}} Project public data releases \citep{Wetzel2023,Grand2024} available on the Globus platform.



\bibliographystyle{mnras}
\bibliography{bibliography} 





\bsp	
\label{lastpage}
\end{document}